\documentclass{PoS}

\title{Nebulosities of the Symbiotic Binary R Aquarii - A Short Review}

\ShortTitle{R Aquarii - A Short Review}

\author{\speaker{Tiina Liimets}\\
Astronomick\'y  \'ustav, Akademie v\v{e}d \v{C}esk\'e republiky, v.v.i., Fri\v{c}ova 298, 251\,65 Ond\v{r}ejov, Czech Republic  \\
Tartu Observatory, University of Tartu, Observatooriumi 1, 61602 T\~oravere, Estonia \\
E-mail: \email{tiina.liimets@asu.cas.cz}}

\author{Romano M. L. Corradi\\
       GRANTECAN, Cuesta de San Jos\'e s/n, E-38712, Bre\~na Baja, La Palma, Spain \\  
        Instituto de Astrof{\'{\i}}sica de Canarias, E-38200 La Laguna, Tenerife, Spain\\
        E-mail: \email{romano.corradi@gtc.iac.es}}
\author{David Jones\\
        Instituto de Astrof{\'{\i}}sica de Canarias, E-38200 La Laguna, Tenerife, Spain\\
        Departamento de Astrof{\'{\i}}sica, Universidad de La Laguna, E-38206 La Laguna, Tenerife, Spain  \\
        E-mail: \email{djones@iac.es}}

\author{Indrek Kolka\\
        Tartu Observatory, University of Tartu, Observatooriumi 1, 61602 T\~oravere, Estonia \\
        E-mail: \email{indrek@to.ee}}

\author{Miguel Santander-Garc{\'{\i}}a\\
        Observatorio Astron\'omico Nacional (OAN-IGN), C/ Alfonso XII, 3, 28014, Madrid, Spain\\
        E-mail: \email{m.santander@oan.es}}

\author{Michael Sidonio\\
         Terroux Observatory, Canberra, Australia\\
        E-mail: \email{m.sidonio@bigpond.com}}

\author{Kristiina Verro\\
        Kapteyn Instituut, Rijksuniversiteit Groningen, Landleven 12, 9747AD Groningen, The Netherlands \\
        E-mail: \email{verro@astro.rug.nl}}

\abstract{
In this proceeding, we present a short review of the fascinating nebulosities of symbiotic binary R Aquarii. The R Aquarii system, comprising the central binary and surrounding nebular material, has been the subject of near-continuous study since its discovery, with a few hundred papers listed in ADS.  As such, it is impossible to provide here the comprehensive review that R Aquarii deserves, instead we chose to focus on the nebulosities -- covering both our own research and other relevant results from the literature.  
}

\FullConference{The Golden Age of Cataclysmic Variables and Related Objects V (GOLDEN2019)\\
		2-7 September 2019\\
		Palermo, Italy}

\begin{document}

\section{Introduction}

R Aquarii (R Aqr) is a symbiotic binary, of orbital period 43.6 years \cite{2009A&A...495..931G}, comprising a mass-losing pulsating Mira (with period 387d) 
and a hot white dwarf (WD). 
It is surrounded by an hour-glass nebula and a curved S-shape jet 
(Figure~\ref{neb}). 
At a kinematic distance of about 200 pc (e.g. \cite{1985A&A...148..274S},  
\cite{2018A&A...612A.118L})
it is the closest known symbiotic system and stellar jet.
Therefore, its environs are very well studied in all wavelengths but still provide many intriguing results and open questions. 
In addition, it is a case study of stellar jets, symbiotic phenomena, mass transfer, as well as mass loss in interacting binaries. 

\section{Optical imaging data} 
We observed R Aqr with the Nordic Optical Telescope exploiting the 
instrument ALFOSC on 18th and 24th of July 2019 within Spanish CAT service time proposal. 
A narrow band filter [O\,{\sc iii}] with a central 
wavelength of 500.7 nm and a FWHM 3 nm (NOT filter \# 90) was used. 
On both dates, one short (10s) and one long (180s) exposure were acquired. 
Images were reduced using standard procedures in 
IRAF\footnote{IRAF is distributed by the National 
Optical Astronomy Observatory, which is operated by the Association 
of Universities for Research in Astronomy (AURA) under cooperative 
agreement with the National Science Foundation.}.

\section{Hour-glass nebula}

\begin{figure}
\centering
\includegraphics[width=0.7\textwidth]{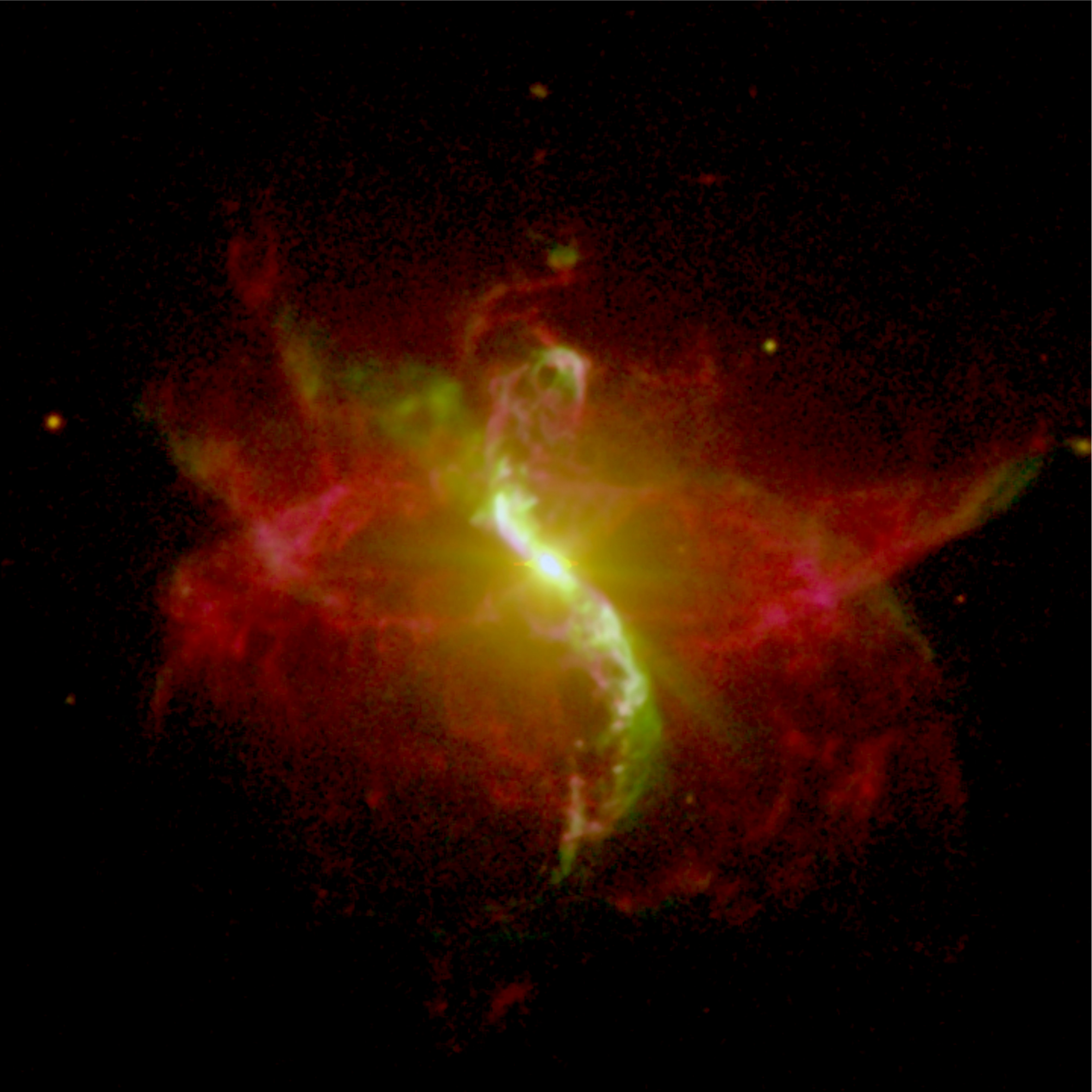}
\caption{ R Aquarii nebulosities. In red the hour-glass nebula, in greenish-yellow 
the jet. North up, East left. 3$'$x3$'$. 
\cite{2018A&A...612A.118L}.}
\label{neb}
\end{figure}

The hour-glass nebula of R Aqr, discovered by \cite{1922PAAS....4..319L} in 1922, has been shown to expand ballistically. 
It has a knotty and filamentary structure in 
the low-ionisation lines (red in Figure~\ref{neb}), such as H$\alpha$, [N\,{\sc ii}], [O\,{\sc i}]. 
In higher ionisation lines, such as [O\,{\sc iii}], the nebula is more diffuse. 
Using the expansion of the nebula several authors, e.g. 
\cite{1944MWOAR..16....1A}, 
\cite{1985A&A...148..274S}, \cite{2018A&A...612A.118L}, 
have derived kinematic ages which all agree that the current age of the R Aqr nebula 
is about 670 years. 
It is believed to have been created by the red giant star in a nova-like event. It was shown 
in \cite{2018A&A...612A.118L} that the nebula does not present noticeable changes during the years 
1991 to 2012. Considering our data from 2019 we can confirm that the nebula has had a stable 
morphology for almost the last 30 years.

\section{New extended features}
Fainter features extending further out beyond the hour-glass nebula, have been 
discovered by \cite{2018A&A...612A.118L}.     
These comprise a thick arc, visible in [O\,{\sc iii}], with a length of 6$'$.4 and a thinner and 
fainter H$\alpha$ arc reaching 2$'$.8 from 
the central star. It is concluded in \cite{2018A&A...612A.118L}
that these features are most likely related to the mass loss from the red 
giant and/or a nova eruption from the white dwarf 
in an earlier evolutionary stage of the system.

\section{Jet}

Unlike the stable bipolar nebula, the evolution of the jet is more complex and irregular. 
The jet has a collimated appearance with a prominent S-shape  
consisting of multiple knotty, filamentary, and diffuse structures 
(Figure~\ref{neb}).

The jet of R Aqr was discovered by \cite{1985A&A...148..274S}. However, \cite{1999ApJ...514..895H} showed 
from the Lowell Observatory photographic plates that the jet was present already as early as 1934.  
The jet presents remarkable brightness, structural, and kinematical 
variations on both large and small 
scales but the overall large-scale S-shape (see Figure~\ref{neb}) has remained the same over the years.  

While the jet of R Aqr is still active (e.g. \cite{2017A&A...602A..53S, 2018A&A...616L...3B}), 
the ages of outer components are up to 300 years old \cite{2018A&A...612A.118L}. 
The jet is formed during the periastron passage of the binary components, when the mass transfer on to the 
white dwarf and surrounding accretion disc is enhanced.

\begin{figure}[t]
\centering
\includegraphics[width=0.42\textwidth]{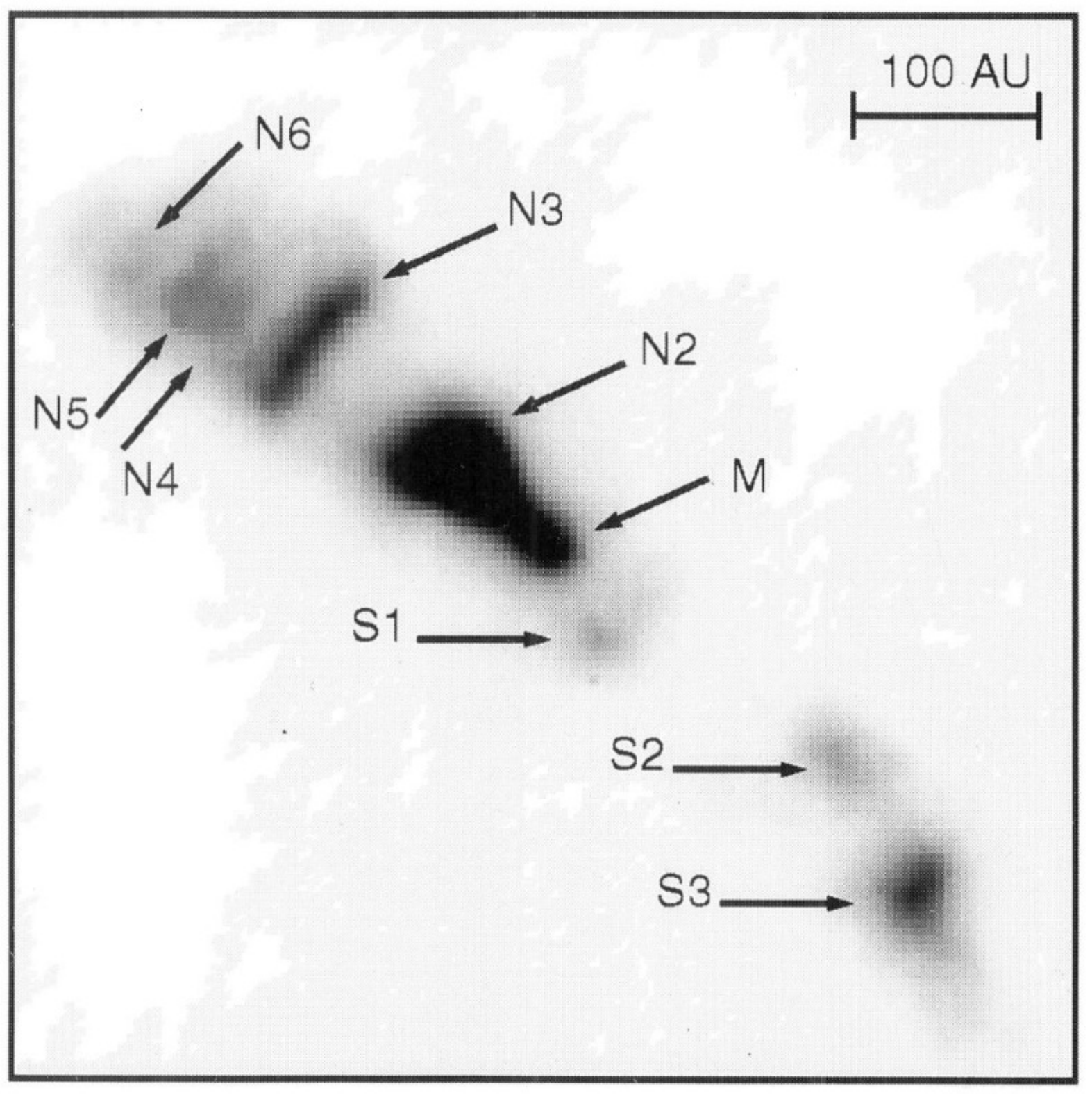}
\includegraphics[width=0.435\textwidth]{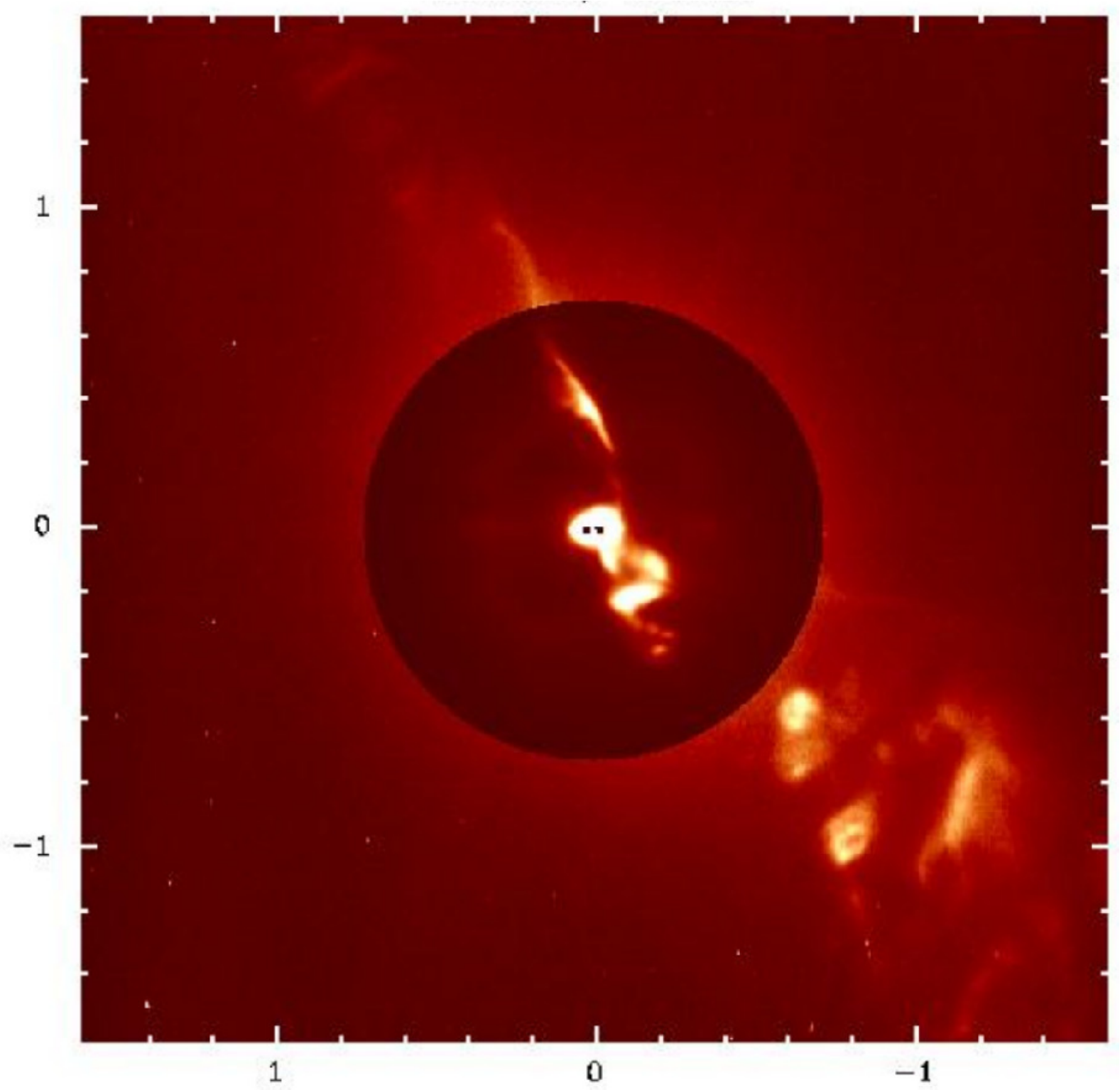}
\caption{\textit{Left:} R Aquarii innermost jet with the HST in UV at 1991. 
Figure from \cite{1994A&A...287..154P}. 
\textit{Right:} VLT+SPHERE H$\alpha$ image from 2014. Figure from \cite{2017A&A...602A..53S}. 
Images have the same FOV, ~3$'$x3$'$. North up, East left. 
}
\label{injet}
\end{figure}
\begin{figure}
\centering
\includegraphics[width=0.44\textwidth]{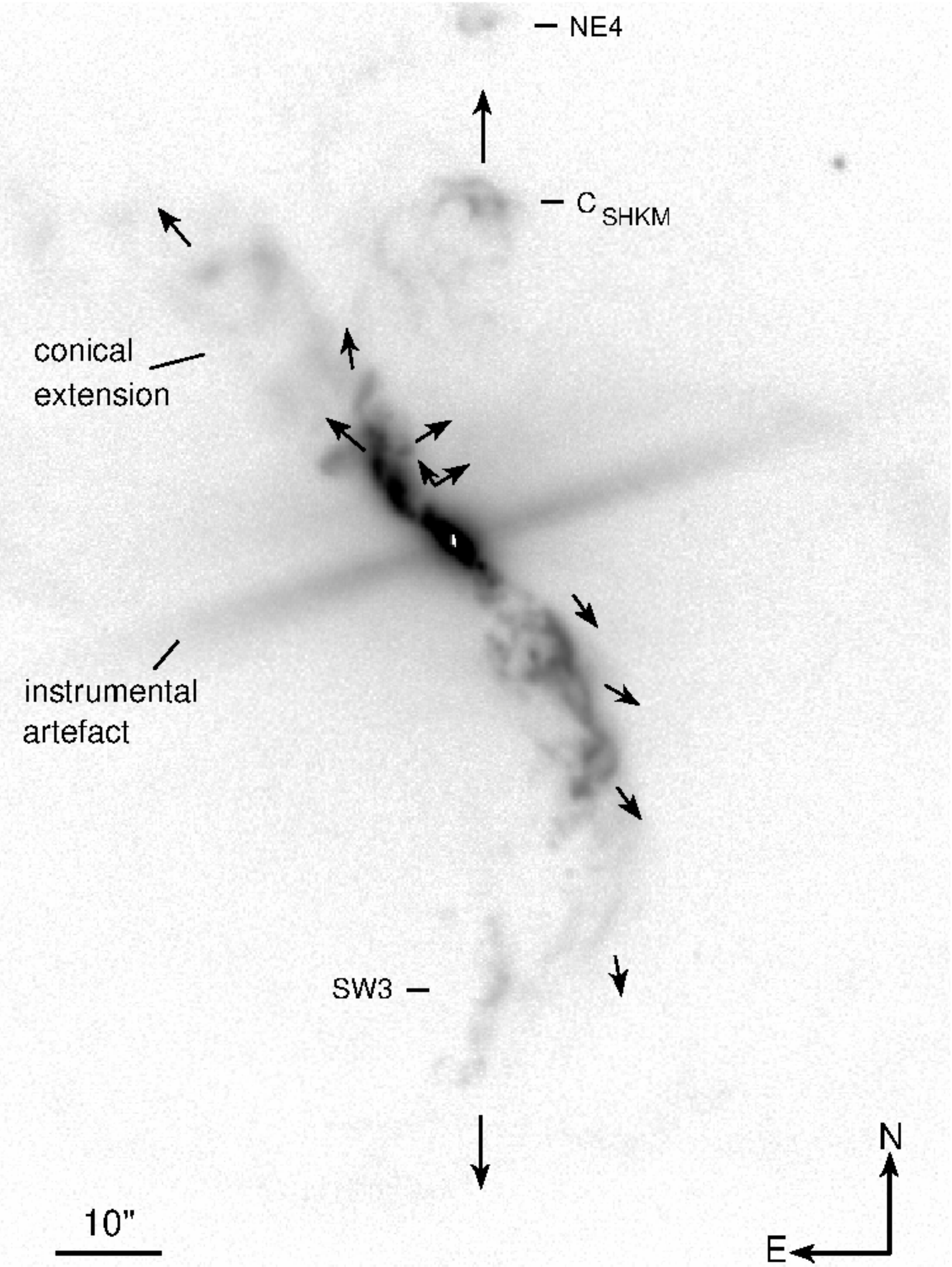}
\caption{Overall expansion pattern of the R Aqr jet. 
Note laterally pointing arrows near the central star in the North-East jet.
Figure depicted from \cite{2018A&A...612A.118L}.}
\label{jetex}
\end{figure}

At earlier times, detailed study of the innermost 5$''$ of the 
jet was only possible with 
high resolution radio data (e.g. \cite{1989ApJ...346..991K}, \cite{2004A&A...424..157M}). 
In 1991, the innermost 3$''$ jet was observed with the Hubble Space Telescope (HST) in the ultraviolet  
(see Figure~\ref{injet} left and \cite{1994A&A...287..154P}).
With the current instruments, it is also possible to probe these central regions in the optical range. 
\cite{2017A&A...602A..53S} observed the R Aqr binary and its jet using the SPHERE instrument on the VLT. 
Their very detailed map of the most central 3$''$ is presented in Figure~\ref{injet} right, 
as a comparison 
with the HST data from 1991. Large differences in brightness and 
morphology between the datasets, taken 23 years apart, 
are evident. In particular, in 1991 the northern jet was considerably brighter than the southern one. 
In addition, in the earlier image the northern filaments were elongated in the North-East direction while 
currently the Northern filaments are stretched out in the North direction, which is the overall jet 
expansion direction (Figure~\ref{jetex}). In the South, the SHPERE data shows a bright ``zig zag'' 
structure which does not seem 
to be present in 1991. These changes are explained by \cite{2017A&A...602A..53S} as resulting 
from the very dynamic nature of the jet. However, as it was shown by \cite{2018A&A...612A.118L}, 
jet features can have a different brightness and morphology depending on the wavelength, 
which could be an additional cause of the mentioned changes in the appearance of the jet 
between the 1991 HST UV and the 2014 VLT+SPHERE H$\alpha$.

\begin{figure}[t]
\centering
\includegraphics[width=0.9\textwidth]{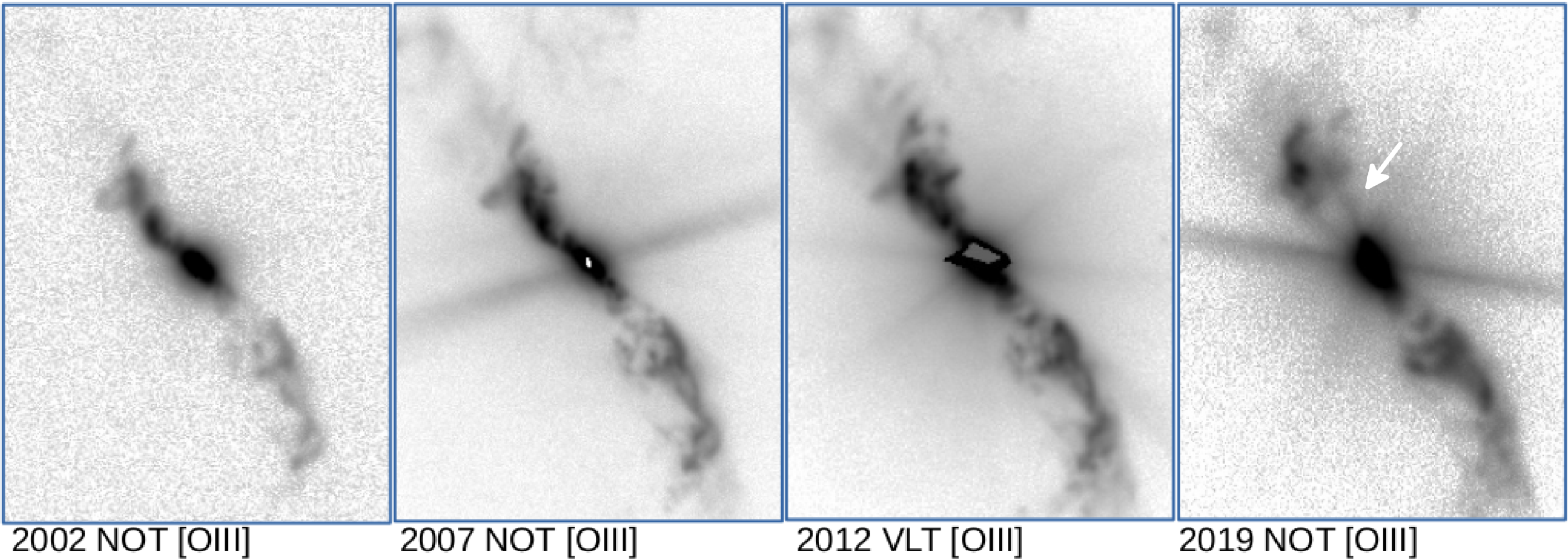}
\caption{R Aquarii jet evolution from 2002 to 2019. 
Diffuse diagonal spikes across the whole FOV, 
emanating from the central star, are instrumental artefacts due to the saturated 
central area.}
\label{jetev}
\end{figure}

Significant changes in the brightness and morphology of the larger scale jet have also been 
seen by \cite{2018A&A...612A.118L} and in the present work. In Figure~\ref{jetev}, we present the evolution 
of the jet on large scales from 2002--2012 (data taken from \cite{2018A&A...612A.118L}), and in 2019 (this work). 
As was described in \cite{2018A&A...612A.118L}, the brightness of the features of the jet changes on 
timescales of about 10 years, in parallel to morphological changes (features tend to get 
elongated in the direction of the overall expansion and eventually separate into two components). 
This is especially true for the North-East jet, while the evolution of South-West jet is more 
uniform. 

Considerable brightness, and possibly also a morphological, changes can be seen between the last
image of \cite{2018A&A...612A.118L} (2012) and our image from 2019: the features closest to 
the central star (highlighted with a white arrow on Figure~\ref{jetev}, or named as A and G by 
\cite{2018A&A...612A.118L}) have completely disappeared 
or at least are no longer detectable on our three minute 
exposure. We would like to point out that all the data 
on Figure~\ref{jetev} have comparable exposure times, indicating that the non-detection is unlikely to be due to a difference in data quality (indeed in 2019, we acquired two epochs roughly one week apart, with the features not visible in either data set) but rather due to the nature of R Aqr and its environs.  Furthermore, HST images from 2013 \cite{2018A&A...612A..77M} 
and 2017 (MAST archive),  show that the differences between 2012 and 2019 have been 
gradually occurring in the interim period. The area highlighted with the white arrow has been getting dimmer 
over the years, until disappearing in 2019. We propose that this large structural change could be related 
to the ongoing period of reduced variability of the Mira in the R Aqr system which started 
at spring 2019 (see AAVSO alert notice 665\footnote{https://www.aavso.org/aavso-alert-notice-665}). 
These low Mira states appear to occur every $\sim$ 44 years, 
synchronised with the orbital period and are associated with the fact that the gas cloud surrounding 
the WD and its accretion disc 
are eclipsing the Mira \cite{1981IBVS.1961....1W}. Considering the refined orbital parameters presented in 
\cite{2018A&A...616L...3B} (see their Figure B.1), it is feasible that the eclipse has indeed started. 
If that is true, then when looking from the Earth, the WD is eclipsing the Mira.  
However, at the same time, when looking from the point of few of the North-East jet, the 
Mira star is eclipsing the light of the WD, which is probably responsible for ionising and illuminating 
the jet in the first place. Therefore, it is not surprising that we see large brightness changes 
that could be associated with the blocking of the ionisation/illumination source. 
The low state of the Mira usually lasts about 8 years, although the passage of the 
WD and the disc should not take much longer than 2 yrs. Therefore, a second explanation to the current 
low state of the Mira is the dimming due to the enhanced mass-loss from the AGB in the direction of 
the WD, which would also be caused by the periastron passage of the system. 
There is an indication of a dust-rich flow of this kind in the ALMA maps \cite{2018A&A...616L...3B}. 
In reality, probably both reasons are playing a role in the current reduced 
brightness variation period of R Aqr. In any case, whatever is the cause, if the brightness of the 
disappearing jet features resumes after 8 years, their fading would almost certainly have been 
related to the current low state of the system.

\cite{2018A&A...612A.118L} presents additional details of the curious nature of R Aqr jet. 
They detect lateral 
fast moving features up to 900 km/s (see Figure~\ref{jetex}), 
which is several times larger than the rest of the 
velocities measured in the jet (e.g. velocities along the line of sight are approximately +- 100 km/s 
\cite{2018A&A...612A.118L}). 
They attribute these fast moving 
features as a changing ionisation/illumination effects rather than a real matter moving.

\cite{2018A&A...612A.118L} also measure the 
northern jet to be mostly red-shifted and southern blue-shifted while previous authors 
(\cite{1985A&A...148..274S}, \cite{1990ApJ...351L..17H}, \cite{1999ApJ...522..297H}) 
have found it to be the opposite. 
This is interpreted as being a consequence of the complex line profiles and the evolution of 
the jet which shows significant 
brightness and morphological changes in the time frame between the 
data in \cite{2018A&A...612A.118L} and 
the previously published radial velocity data. Moreover, the data used in \cite{2018A&A...612A.118L} 
has a much higher spectral and spatial resolution than any other dataset published before.

Considering all the above mentioned intriguing results and the jet's kinematical behaviour, it is worth emphasising 
that the overall large scale expansion pattern of the jet can be still considered ballistic 
(Figure~\ref{jetex} and \cite{2018A&A...612A.118L}).

\section{Conclusions}

We have presented a short review of the R Aqr nebulosities together with 
the surprising morphological and brightness changes observed during the last 7 years. 
The latter we believe is associated with the current possible eclipse of the Mira 
by the WD and its accretion disc or by the dimming due to the enhances mass-loss from the AGB, 
both caused by the periastron passage. 
From all this it is clear that 
the evolution of the R Aqr jet cannot be described by purely radial expansion. 
A combination of physical matter moving, together with changing ionisation, illumination, 
shocks, and precession has to be taken into account. However, which processes 
from those listed is more important in the evolution of individual features 
remains an open question.  

\acknowledgments
Based on observations made with the Nordic Optical Telescope, operated by 
the Nordic Optical Telescope Scientific Association at the Observatorio del 
Roque de los Muchachos, La Palma, Spain, of the Instituto de Astrofisica 
de Canarias. The data presented here were obtained with ALFOSC, 
which is provided by the Instituto de Astrofisica de Andalucia (IAA) under 
a joint agreement with the University of Copenhagen and NOTSA. 
DJ acknowledges support from the State Research Agency (AEI) 
of the Spanish Ministry of Science, Innovation and Universities (MCIU) and 
the European Regional Development Fund (FEDER) under grant AYA2017-83383-P.
IK acknowledges the support of the Estonian Ministry for
Education and Science under grant IUT40-1. 
TL acknowledges financial support from GA\v{C}R (grant number 17-02337S). 
The Astronomical Institute Ond\v{r}ejov is supported by the project 
RVO:67985815.

\bibliographystyle{JHEP}
\bibliography{literature}

%

\bigskip
\bigskip
\noindent {\bf DISCUSSION}

\bigskip
\noindent {\bf VALENTIN BUJARRABAL:} you mentioned an East-West "bipolar nebula", different from 
the jets you are mainly interested in. But, in view of its velocity field and structure, 
I would say that it is probably a ring in expansion. Possibly, it defines the plane 
of symmetry of the whole extended nebula. Do you agree?

\bigskip
\noindent {\bf TIINA LIIMETS:} yes

\end{document}